%% file: main.tex
\documentclass[sigplan,10pt]{acmart}
\makeatletter                   
\def\mdseries@tt{m}             
\makeatother                    
\usepackage[draft=true]{minted}
\usepackage{hyperref}
\usepackage{enumitem}
\usepackage{subcaption} 
\renewcommand\footnotetextcopyrightpermission[1]{}
\setcopyright{none}
\acmConference{}
\acmYear{}
\acmISBN{} 
\acmDOI{} 
\startPage{1}

\begin{document}

%

\title{\emph{Kvik}: A task based middleware with composable scheduling policies}

\author{Saurabh Raje}
\email{saraje01@in.ibm.com}
\affiliation{%
  \institution{IBM Research}
  \city{New Delhi}
  \country{India}
}
\author{Fr\'ed\'eric Wagner}
\email{frederic.wagner@univ-grenoble-alpes.fr}
\orcid{0000-0002-5261-8748}
\affiliation{
    \institution{Univ. Grenoble Alpes, CNRS, Inria, Grenoble INP*, LIG}
    \city{Grenoble}
    \postcode{38000}
    \country{France}
}


\begin{abstract}
  In this paper we present \emph{Kvik}: an implementation
    of a task-based "middleware"
    for shared memory parallel programming in the \emph{Rust} language
    built on top of the \emph{Rayon} library.
  We devise a system allowing several task-splitting schedulers to be finely tuned by the
  end users. Among these, we propose an implementation of an \emph{adaptive}
  scheduler reducing tasks creations (splits) to bare minimum by linking tasks splitting
  to steal requests. Another important scheduler that allows turning computations
  into sequences of parallel operations is described. This operator proves itself particularly
  useful for interruptible computations. We exhibit different code examples
  well suited for different types of schedulers.
  We conclude our work with a set of benchmarks making heavy use of composability. In particular
  we present a parallel stable sort implementation with up to 1.5x more speedup when
  compared to the state-of-the-art parallel sorting implementation.
\end{abstract}

\maketitle

\section{Introduction}

Parallel programming has long been one of the most difficult form of programming.
Accelerating computations on a parallel architecture demands a complete understanding
of an entire stack of abstractions - from parallel algorithm design and analysis,
down to the processor and memory configuration.

To lessen this problem task-based parallel programming builds an accessible bridge between parallel algorithm design and its implementation.
A taxonomy of task-based parallel programming tools is presented in \cite{taxonomytaskbased}.
The user is discharged of scheduling issues deferred to the run-time of task management middleware.
For example, SPMD parallelism can be trivially scheduled with no dependencies.
Unfortunately, this design trivializes the question of task splitting. Consider a recursive divide and conquer algorithm that uses the fork-join model for the tasks. Each call can be potentially run in parallel, and hence is mapped to a task.
There is a stop condition for this recursion, and that is where the task-splitting stops.
As a result, the algorithm implementation and task splitting are very tightly coupled.
Any attempts to tweak performance using different task splitting strategies requires the programmer to dig into
the implementation of the algorithm itself.

The \emph{Rust} language is getting a lot of attention these days due to its unique features related
to safety~\cite{balasubramanian2017system}, that also make it a good candidate for parallel programming.
The language benefits from a very strong memory model that has ownership and borrowing as its first class concepts.
The compiler disallows multiple "overlapping" mutable aliases on the same object. This functionality is called the "borrow checker" and is notorious for its steep learning curve. This one change however, makes parallel programming inherently safe in \emph{Rust}.
\emph{Rust} also has a powerful set of generics ("traits") for concurrency that govern which types
can be shared (\href{https://doc.rust-lang.org/std/marker/trait.Sync.html}{\emph{Sync}})
or moved (\href{https://doc.rust-lang.org/std/marker/trait.Send.html}{\emph{Send}}) between threads.
For example, \emph{Rust} provides two different types for reference counting: \emph{Rc} and \emph{Arc},
the latter one using atomic counters. The type system will detect at compile time the
sharing of \emph{Rc} type between two different threads, since \emph{Rc} does not implement \emph{Sync}.

Finally \emph{Rust} provides a functional-style API that is well suited for expressing
clean parallel code. In particular the \emph{Rayon}~\cite{rayon} library already allows many users
(more than 7,000,000 downloads as of July 2020) to write safe and elegant parallel code in a functional manner, using task splitting and stealing to parallelize the computation.
Given its functional nature, the notion of tasks and dependencies is exposed organically.
The contributions of this paper are:
\begin{enumerate}
        \item We introduce \emph{Kvik}, a prototype implementation of configurable schedulers that exposes a functional API for algorithm design and can finely tune task splitting.
        \item We propose a modular mechanism for tweaking the behavior of schedulers.
            This \emph{Adaptor} API is accessible to the end users and allows them to modify the internal scheduling policies.
            It also allows for easy composition of multiple policies.
        \item We propose a new API for abstracting divide and conquer computations.
            This API allows to delegate decisions regarding the splitting of tasks to the middleware.
        \item We demonstrate up to 1.5x more speedup on a parallel stable sort (over the state of the art) tuned using a variety of task splitting schedules, and packaged with \emph{Kvik}. Additional benchmarks demonstrate the performances
            of different adaptors and schedulers.
\end{enumerate}

We provide some \emph{Rust} code in different places through this paper.
The code is always very short and should be clear even for readers who don't know the language.
We also use links to source codes for further details.

The paper starts off in Section~\ref{sec:middlewares} by introducing several
features of current task-based middlewares upon which we rely. We continue by presenting
the inner workings of our library along with several code examples in Section~\ref{sec:Kvik}.
In particular we develop several schedules and many adaptors that control task division.
These developments give the programmer a fine-grained control over the behavior of their parallel algorithm in very few lines of code.
Section~\ref{sec:expes} presents some experimental results on a set of different algorithms.
We show examples relying on different scheduling strategies for their
optimal executions.
Finally we conclude on our work in Section~\ref{sec:conclusion}.

\section{Related Work}
\label{sec:middlewares}

In recent years we have seen many new developments
of task-based middlewares:
\emph{Cilk}~\cite{blumofe1996cilk},
\emph{TBB}~\cite{kukanov2007foundations},
\emph{OpenMP}~\cite{dagum1998openmp},
\emph{XKaapi}~\cite{Gautier07kaapi:a},
\emph{TPL}~\cite{tpldotnet} for the .NET framework,
 and \emph{Rayon}~\cite{rayon} for Rust.
Task based parallel programming has also been applied to distributed memory models \cite{ductteip}.
A very recent work \emph{Cpp-Taskflow}~\cite{cpptaskflow}, parallelizes code through task-graphs given by the programmer.
Similarly, SMPSs~\cite{smps} uses directives to allow the programmer to describe tasks. The runtime then dynamically makes a dependency graph.
Legion~\cite{legion} and Chapel~\cite{chapel} introduced support for data partitioning and task based programming at a language level.
HPX~\cite{hpx} is a runtime-environment that combines task-based programming with a Global Address Space.

We should note that of all the libraries listed in this section, only
\emph{Rayon} provides a programming interface in a functional programming style.
To the best of our knowledge, exposing task splitting schedules to the programmer and allowing composability of such schedules
is a novel contribution of \emph{Kvik} compared to the previous state of the art.
Dependencies can be expressed organically using a functional programming style.
Finally, the memory safety guarantees afforded by Rust make \emph{Heisenbugs} extremely uncommon, at no cost to performance.

Load balancing in task based programming is usually achieved through a \emph{work stealing} engine.
Available tasks are distributed among threads and any thread becoming idle
will seek additional work from others. Traditionally this choice is random but
different other policies have been developed \cite{adws}.
Work stealing is well known~\cite{blumofe1999scheduling} for its theoretical guarantees, bounding the number
of steal requests by $O(p\times D)$ where $p$ is the number of threads and $D$
the depth of the algorithm. The cost of maintaining the lists of tasks to be stolen from is studied in \cite{workstealingdenis}.

We now zoom-in on some middlewares for the comprehension of our work.

\subsection{TBB}
\label{sec:TBB}

Intel TBB is an important task-based middleware with many applications.
The overheads of synchronization and stealing in TBB have been well studied in \cite{contrerasTBB}.
In this section we are going to zoom on its grain size determination mechanism~\cite{4536188}
which is also used in \emph{Rayon}.
This mechanism is important because it allows to dynamically tie the tasks creations
to work-stealing. The benefits are two-fold:
\begin{itemize}
    \item Better performance due to less task creations, divisions and reductions;
    \item No need for the end user to specify a sequential fallback size.
\end{itemize}

The task splitting policy works as follows:
\begin{enumerate}
    \item Start with an initial task and associate it to a counter (usually a multiple of the number $p$ of threads)
    \item When the task gets divided, divide counter by $2$.
    \item If it reaches 1, stop creating new tasks and compute what is left to do in sequential.
    \item When a task gets stolen however, reset the counter to a higher value in order to enable the creation of new tasks.
\end{enumerate}

This policy is quite well designed. If the work is balanced, threads synchronized and their number
a power of two, we end up creating $O(p)$ tasks. In other cases the number
of tasks created might be higher. For example imagine three threads. Since the initial
task gets divided in two, we create four tasks to feed all threads. Three tasks get
completed in parallel and we are left with a single task for three threads.
The whole process then repeats itself until tasks cannot be divided anymore.
For an input size of $n$ this gets repeated $O(\log(n))$ times if all divisions are always possible (which corresponds
here to the depth). This is still much better than a naive $\Omega(n)$ tasks creations.
We have been able to confirm this expected behavior with our middleware through various experiments.

In \emph{Kvik}, we provide this task splitting policy to the programmer.
It is called as the \emph{thief\_splitting} strategy.

\subsection{Xkaapi} 
\label{sec:kaapi}

\emph{Kaapi}~\cite{Gautier07kaapi:a} (now renamed as \emph{Xkaapi}) is an experimental
task-based middleware.
Efforts were focused on minimizing overheads,
in particular overheads related to task creation.

Of particular interest to us is the ability for \emph{Xkaapi} to execute
adaptive parallel algorithms~\cite{adapt}.
The main idea is to delay task division as much as
possible to limit task creation overhead.
This is achieved by linking task division to steal requests.
If there is no steal request because all threads are busy,
no task division occurs and all computations take place in a single task.

While this idea is compelling its implementation is rather difficult:
it requires a way to interrupt a running task when a steal requests occurs.
\emph{Xkaapi} achieves this through nested loops:
\begin{itemize}
    \item an innermost \emph{nano}-loop performs uninterrupted sequential computations
    \item an outer \emph{micro}-loop that:
        \begin{itemize}
            \item executes the \emph{nano}-loop for some iterations
            \item checks if a steal request is received, and then divides the task
        \end{itemize}
\end{itemize}

The idea of the \emph{nano}-loop is that steal request can not be checked for
at every iteration since checking disrupts the flow of instructions, and involves \emph{atomics}.
Loop nesting amortizes this cost by checking less often.

Now one question remains: what block sizes should be picked for the \emph{nano}-loop ?
While a small size generates some overheads, a large size means that
stealers can get blocked for a longer time.
Of course for a given computation, an optimal size can be found after some benchmark runs.
However, this task is context dependent and needs programmer intervention.

A generic solution is to take a geometric series as a sequence of sizes.
For example, the middleware can start at size $1$ and double the size at each un-stolen \emph{micro}-loop.
This way if $n$ is the total number of loops to execute, there can not be more than
$\lceil \log(n) \rceil$ \emph{micro}-loops. The amount of time spent in wait by a stealer
cannot be more than the actual time spent working usefully.
At each steal, the size is reset to $1$.

\subsection{Rayon}
\label{sec:rayon}

\emph{Rayon}~\cite{rayon} is a recent work-stealing middleware developed
in the \emph{Rust} programming language. It benefits from two key aspects of \emph{Rust}:
enhanced security through rust's security mechanisms (borrow checker and type system)
and a functional programming style. Since it serves as the base for our work we detail
some of its inner mechanisms.

The base operation in \emph{Rayon} is the \mintinline{rust}{join()} function which:
\begin{itemize}
    \item Takes as arguments two closures to be run in parallel.
    \item Creates a task for the second closure and executes the first one immediately.
    \item Blocks the thread (goes stealing) until both closures are executed.
\end{itemize}
In \emph{Kvik}, we directly use this function for executing the tasks created.
It hence allows us to reuse the work stealing engine of \emph{Rayon} as-is.
The task splitting schedule in \emph{Rayon} is the same as TBB.
The main feature of \emph{Rayon} is that it provides \emph{parallel iterators}.
We now describe how parallel iterators are implemented.

\subsubsection{Parallel iterators}
\label{subsec:pariter}

A \emph{ParallelIterator} trait in \emph{Rayon} extends the concept of sequential iterators
to parallel iterations. We can for example compute $\sum_{i=0}^9 f(i)$ with:

\begin{minted}{rust}
    (0..10).into_par_iter().map(f).sum()
\end{minted}

For many algorithms turning to parallel code is as simple as switching from \mintinline{rust}{.into_iter()}
to \mintinline{rust}{.into_par_iter()}.

As a usage example for \emph{Rayon},
let's take as input a vector $v$ of integers and compute a new vector
containing all even elements of the input in parallel.
This operation is known to be not trivial to program. The fact the elements
are filtered out makes it difficult to move elements to their final positions
in parallel.

\begin{minted}{rust}
v.into_par_iter() // par iter on integers
 .filter(|&e| e % 2 == 0) // only even integers
 .fold(Vec::new, |mut v, e| {
     v.push(e); // each task produces one vector
     v
 })
 .map(|v| once(v).collect::<LinkedList<_>>())
    // List containing 1 vector
 .reduce(LinkedList::new, |mut l1, mut l2| {
     l1.append(&mut l2);
     // concatenate lists of vectors in parallel
     l1
 })
 .into_iter() // loop on all vectors sequentially
 .flatten()
 .collect::<Vec<_>>(); // into one vector of ints
\end{minted}

The computation starts with the filtered iterator on the input.
This iterator is going to be divided into smaller chunks dynamically by \emph{Rayon}'s scheduler.
Each division will result in a call to the \mintinline{rust}{join()} to create new tasks.
Once the division stops, a sequential \mintinline{rust}{fold()} operation is applied on the chunk.
This produces a small vector that contains all the elements of the chunk.
As a result, there exists one vector (with only even elements) corresponding to each task that did not divide further.
Every vector is then mapped into a linked list.
A parallel reduction then concatenates the multiple lists into one single list.
To complete the algorithm, the list of vectors is turned back into a sequential iterator
and flattened into a single vector.
Note that the programmer can not specify how small each task should be in this example.
Hence, while it is quite easy to perform complex computations, task splitting is invisible to the programmer.

\subsubsection{Producers}
Implementing the entire \emph{ParallelIterator}-based pipeline is done through divide and conquer as follows:

\begin{enumerate}
    \item We start from the initial data and divide it into two parts recursively, forming a division tree.
    \item When leaves are reached data is folded sequentially.
    \item The results are reduced two-by-two forming a reduction tree symmetrical to the division tree.
\end{enumerate}

However \emph{ParallelIterator}s, like all iterators, are lazy and are only holding data and functions
while waiting for computations to start. The operation which will launch all computations is
the \mintinline{rust}{reduce()}.
It takes a function generating identity elements, and a function that turns two elements into one.

Most types that implement \emph{ParallelIterator} cannot be divided into two.
Like the \emph{Map} for example which is obtained by applying
a function on each element of an underlying base iterator. A \emph{Map}
\href{https://docs.rs/rayon/1.3.1/src/rayon/iter/map.rs.html#15-18}{structure} contains two
fields: the base iterator and the function to apply.
Due to \emph{Rust}'s very specific memory model that requires
data to have exactly one owner, the function (can be a closure)
cannot be owned by two different structures.
The simple solution to this problem is to take it by reference.
This requires a
\href{https://docs.rs/rayon/1.3.1/src/rayon/iter/map.rs.html#113-116}{different structure}
which still has two fields but one field is now a reference of the real function.
We can abstract over these types by introducing a new trait: the \emph{Producer} trait.

A type that implements \emph{Producer} can not only be divided into two pieces,
it can also be iterated over as it produces some items.
This is why it is called a \emph{Producer}.
While types that implement \emph{ParallelIterator} simply hold data and/or computations,
types that implement \emph{Producer} really carry them out.
Hence, the \emph{ParallelIterator} is turned into a \emph{Producer},
and the above divide and conquer algorithm is run on the \emph{Producer}.

\section{Kvik}
\label{sec:Kvik}

In this section we present \emph{Kvik}, our prototype for an alternative implementation
of \emph{Rayon}'s \emph{ParallelIterator}. \emph{Kvik} stands for \emph{Kaarya VIbhaajaK}, which is Hindi phonetic for "Task Splitter". We try to fulfill several goals:
\begin{itemize}
    \item be faster
    \item allow more expressivity
    \item allow the end user to define and control task splitting policies
\end{itemize}

\emph{Kvik} allows the end user to express very complex parallel algorithms in few lines
of code with excellent performances and interchangeable schedulers.

We start by introducing in Section~\ref{sec:divisible} the
\emph{Divisible} trait which is the most fundamental abstraction
we use. We then present Section~\ref{sec:join} a basic
scheduler for tasks creations. Section~\ref{sec:control} we
show how to modify on demand the scheduler behavior through \emph{adaptors}.
We show how to abstract other more generic
divide and conquer schemes in Section~\ref{sec:wrap_iter}.
We then provide two different performance enhancing schedulers.
Section~\ref{sec:try_reduce} we provide a scheduler
using a sequence of parallel blocks and Section~\ref{sec:adaptive}
an adaptive scheduler linking tasks creations decisions
to steal requests.
Finally we
conclude Section~\ref{sec:mergesort} with an elegant
parallel merge sort algorithm putting all introduced
features into use.

\subsection{The \emph{Divisible} trait}
\label{sec:divisible}

We introduce a new trait: \emph{Divisible}, which requires the following functions to be implemented:
\begin{itemize}
    \setlength{\itemindent}{-1.2em}
    \item \mintinline{rust}{fn should_be_divided(&self) -> bool;}
    \item \mintinline{rust}{fn divide(self) -> (Self, Self);}
    \item \mintinline{rust}{fn divide_at(self, index: usize) -> (Self, Self);}
\end{itemize}

The \emph{divide} method takes one object and divides it into two objects containing
the left and right parts of underlying data. Left and right parts are expected to be approximately
balanced but this is not always the case.
The \emph{divide\_at} method takes an additional index at which to cut the object into two.
The left part being approximately of size \emph{index}. This is an important method since
some algorithms rely on uneven divisions to be efficient (see section~\ref{sec:try_reduce}).
Finally the \emph{should\_be\_divided} method takes the object by reference and asks it
whether or not it should be further divided. This method
allows to delegate task creations decisions to user-space.

Once we have this, we can define \emph{Producers} (in \emph{Kvik}) as types
which are both sequential \emph{Iterators} and \emph{Divisible}.

\subsection{Join scheduler}
\label{sec:join}

We now describe the implementation of the scheduler that can create tasks,
facilitate work stealing, and finally reduce the results from the tasks that have terminated.
The most naive implementation for such a scheduler is based on the fork-join model.
It makes use of the \mintinline{rust}{join()} from the Rayon library and is hence called \emph{join scheduler}.

\begin{minted}{rust}
fn schedule_join(&self, producer: P, reducer: &R)
    -> P::Item {
    if producer.should_be_divided() {
        let (left, right) = producer.divide();
        let (left_r, right_r) = rayon::join(
            || schedule_join(left, reducer),
            || schedule_join(right, reducer),
        );
        reducer.reduce(left_r, right_r)
    } else {
        reducer.fold(producer)
    }
}
\end{minted}

The decision to divide and create tasks is delegated to the producer with \mintinline{rust}{producer.should_be_divided()} (A \emph{Producer} always implements \emph{Divisible}).
The tasks are created using the \mintinline{rust}{join()} from \mintinline{rust}{Rayon} (see Section~\ref{sec:rayon}).
This uses the existing work stealing implementation from \emph{Rayon}.
Finally, the results from the two tasks are reduced into one.

We also provide a superior variant called \emph{schedule\_depjoin} which allows
the reduction to be executed without wait
by the last thread to finish one of the two parallel operations.
In contrast, as per the standard \emph{rayon::join}, if the thread executing the left task finishes first,
it has to wait for the right result.
It hence starts stealing tasks.
If it gets a task, it will not
initiate the reduction until the stolen task is complete.

\subsection{Adaptors to control task splitting}
\label{sec:control}

The \mintinline{rust}{should_be_divided} method enables us to write many interesting adaptors
controlling the division process in different ways.
Adaptors can be nested into each other and provide
a high degree of composability.

For example, by default, basic producers
(on slices or ranges) divide themselves until a size of $1$ is reached.
It is pretty common however to stop the creation of parallel tasks way before reaching
the smallest size. It is easy to write an adaptor which contains
an underlying producer, a depth limit
and a division counter. When divided both left and right subparts increase the division counter.
If the counter reaches the limit then \emph{should\_be\_divided} returns false. This adaptor
is in fact available as \emph{bound\_depth} in \emph{Kvik}.
It can be used easily like \mintinline{rust}{(0..1000).into_par_iter().bound_depth(3).sum()}.
In this example the parallel range \emph{Producer} is turned into a \emph{BoundDepth} \emph{Producer}.
When the sum reduction is scheduled, the initial range is divided recursively $3$ times until
$8$ tasks are formed.
Note that all this code is written in user space.
We also provide the following useful adaptors:

\begin{itemize}
    \item \emph{even\_levels}: enforces all leaves to be on an even depth level.
        Implemented by flipping a boolean every time it is divided.
    \item \emph{force\_depth}: enforces the division tree to be a complete tree for at least the given depth.
    \item \emph{size\_limit}: stops divisions when the underlying producer is of given size.
    \item \emph{cap}: counts the active number of tasks and refuses division when the number reaches a threshold.
        This also decrements the counter as the tasks finish.
    \item \emph{join\_context\_policy}: divides tasks upto a given depth.
        Left tasks are always divided and right tasks only if stolen.
\end{itemize}

Finally, there is a \emph{thief\_splitting} adaptor that allows to re-implement
\emph{Rayon} and \emph{TBB}'s mechanism for controlling task divisions:

\begin{enumerate}
    \item start with a counter and the thread ID of the thread that created the task
    \item when divided the counter gets decreased by one, children get a copy of parent's thread ID
    \item if the counter reaches zero, \emph{should\_be\_divided} returns false unless the parent's thread ID is not the same as child's thread ID
    \item if task is stolen, the counter is reset to its initial value
\end{enumerate}
In the \emph{Rayon} library this counter has an initial value equal to $\log p + 1$ for $p$ threads
in order to force the creation of $2p$ tasks.
Here, this initial value can be given by the programmer.

Adaptors strongly control the scheduler
and can be turned on or off by the end user.
This allows an easy way to compare
different scheduling policies
and to tune algorithms to the best one.

\subsection{Parallelizing divide and conquer}
\label{sec:wrap_iter}

The \emph{Divisible} trait also contains a pre-implemented funtion \mintinline{rust}{wrap_iter()}
that allows the programmer to easily write parallel divide and conquer algorithms.

Let's take a classical divide and conquer \emph{maximum subarray sum} problem.
The input is recursively divided in two, looking for maximum sums in the left
and right pieces.
On return we then search for the maximum sum touching the midpoint
before returning the max of the left, right and middle sums.

This can be parallelized naively, by turning recursive calls into parallel
recursive calls just using \mintinline{rust}{rayon::join()}.
However, the sequential fallback size would then
have to be manually written inside an \emph{if} condition.

A more generic option is to use \mintinline{rust}{wrap_iter()} as follows:

\begin{minted}{rust}
fn max_sum_par(slice: &[i32]) -> i32 {
slice
    .wrap_iter()
    .map(|s| (s, max_sum_seq(s)))
    .thief_splitting(4)
    .reduce_with(|(left, l_sum), (right, r_sum)| {
        let mid_sum = middle_sum(left, right);
        (
            fuse_slices(left, right),
            l_sum.max(r_sum).max(mid_sum),
        )
    })
}
\end{minted}

In this example, \emph{wrap\_iter} will turn the \emph{slice} into a parallel iterator
on sub-\emph{slice}s. These subslices are then mapped to compute the max sum sequentially
on each one (as a side note, this can be a faster algorithm).
By default, slice would divide until a size $1$. Using the \emph{thief\_splitting} adaptor,
the scheduling policy is changed to restrict tasks creation.
Individual results are finally fused back together (in parallel) by computing the middle-part sum
and comparing all sums at each reduction step.
This example can be extended to a more generic use-case where the input is not a \emph{slice},
but any type that implements the \emph{Divisible} trait.
Furthermore, instead of \emph{thief\_splitting} any adaptor(s) can be put after the \emph{map}.

\subsection{Scheduling sequences of parallel operations}
\label{sec:try_reduce}

There is interesting class of problems which are easy to parallelize,
but a naive implementation increases the work and doesn't scale well.
For example, the \emph{find\_first} that applies a function $f$ on all
elements of an iterator and returns the first element $e_i$ ($i$ minimal)
for which $f(e_i)$ is true.
For $P$ threads running, it is tempting to simply partition the input of size $N$
into ${P}$ pieces and give one to each thread.
However, if the minimal $i$ lies at $\frac{N}{P} - 1$, this gives no speedup!
To solve this problem, we introduce the \emph{by\_blocks} adaptor.

Instead of parallelizing all computations over the \emph{Producer}, the \emph{by\_blocks} will
divide the producer into blocks of growing sizes, and advance
{\bf sequentially} over the blocks.
Each block is scheduled in parallel using all available threads.
This scheme is implemented using the \emph{divide\_at} method of the \emph{Divisible}
trait that cuts the \emph{Producer} at a given point.


We need however, to choose adequate block sizes.
A good solution is again to use a geometric series.
For example we take the number of threads ($P$) for the initial size,
process the first block in parallel and double the size,
process the second block in parallel and double again the size and so on.
Using this series, the number of blocks is logarithmic in input size.
Since useless computations can only take place in the last block,
and this block's size is approximately equal to the sum of sizes of
all the blocks processed before, we have a bound on the amount of useless work.
It cannot be more than one half of the total work.
Changing the increase factor will of course adjust this ratio to whatever the user likes.
In contrast, for the naive partitioning, up to $\frac{N}{P} * {P-1}$
can be useless.

\begin{figure}[htbp]
    \begin{center}
        \includegraphics[width=7cm]{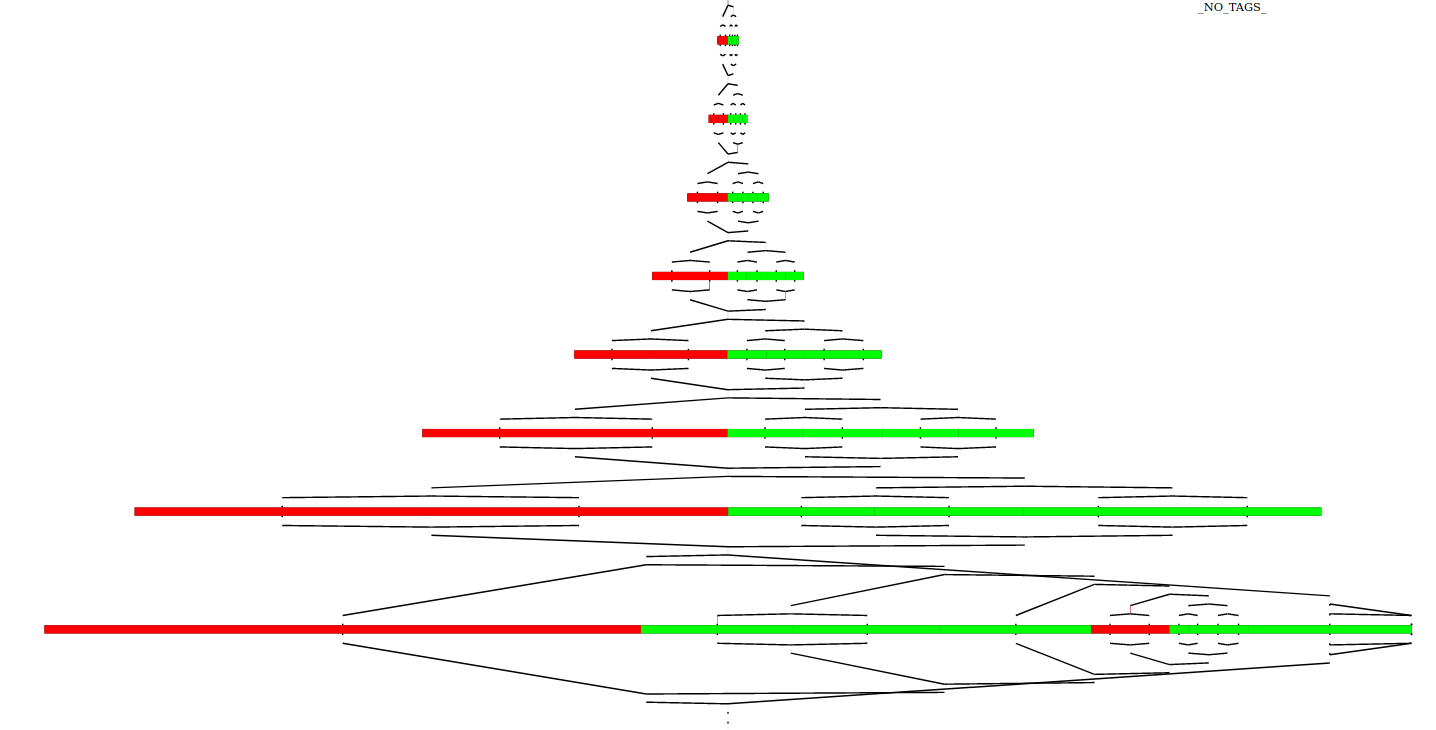}
    \end{center}
    \caption{Executing \emph{find\_first} with 2 threads (one color per thread). \emph{thief\_splitting} adaptor inside each block.
    The sequence of blocks of sizes growing exponentially is visible top to bottom.}
    \label{fig:find_first}
\end{figure}

Figure~\ref{fig:find_first} displays this schedule in action.

\subsection{Adaptive scheduling}
\label{sec:adaptive}

We provide the adaptive schedule from Section~\ref{sec:kaapi} for use in \emph{Kvik}.
In order to achieve this we require a new method in the \emph{Producer} trait: \emph{partial\_fold}.

\begin{minted}{rust}
fn partial_fold<B, F>(&mut self, init: B,
    fold_op: F, limit: usize) -> B
where
    B: Send,
    F: Fn(B, Self::Item) -> B;
\end{minted}

This is similar to a fold operation except that it needs a \emph{limit} on the number of iterations to run.
It will also borrow mutably the \emph{Producer}, making it usable after a fold.
The \emph{partial\_fold} hence replaces the nano-loop of the adaptive scheduling algorithm.

The scheduler code is much more complex (60 lines) than the simple join scheduler since
we need to re-implement the whole mechanism of doing sequential work, checking for
steal requests and dividing the producer if stolen. However, this complexity
is hidden from the programmer, who just need to call the \emph{adaptive}
adaptor to switch to adaptive scheduling.

The benefits are:
\begin{itemize}
    \item less tasks creations (number of successful steals + 1)
    \item less divisions and reductions
    \item on division the \emph{remaining work} is divided in two, so the division is fairer
\end{itemize}

\subsubsection{Adaptive divide and conquer}
While the adaptive schedule is nice, it cannot be used
in conjunction with \emph{wrap\_iter} from Section~\ref{sec:wrap_iter} because types which are only \emph{Divisible} and not \emph{Iterator}s do not provide any way to fold partially.
For such cases the end user needs to provide additional information
on the computations executed in the nano-loop. We provide
therefore a \mintinline{rust}{work} method.
\label{sec:work}
This method offers stateful nano-loops for \emph{Divisible} states, and needs two arguments, a closure $C$ and the initial state.
The closure $C$ (written by the user) should borrow (mutably) the \emph{State} and take-in
an integer $I$.
With the closure, the user describes how to work upon a given state for $I$ iterations.

This dovetails nicely with the \emph{partial\_fold} above,
where the closure $C$ is called (instead of the \emph{fold\_op}) with the \emph{limit} as $I$.

\mintinline{rust}{work()} is particularly useful to implement algorithms
which are hard to write in a pure functional programming style.
A good example is graph traversals, for which the state can consist of the stack or queue,
and a reference to a shared set of visited nodes.

\subsection{Parallel merge sorts}
\label{sec:mergesort}
In this section we present a parallel merge sort algorithm
which takes advantage of most of \emph{Kvik}'s features.
It is a good example expliciting why composition matters.

We start with a tuple of two mutable slices: the input slice (formed over the input Vector) and a temporary buffer of the same size.
Since mutable slices are \emph{Divisible}, a tuple of mutable slices is also \emph{Divisible}
(the division splits each slice into two, and yields two tuples with the left and the right splits of both slices).
We then call \emph{wrap\_iter} (see \ref{sec:wrap_iter}) on the tuple to get a \emph{ParallelIterator}.
Note that the item yielded by this ParallelIterator is a tuple of sub-slices.

With this, we can write the first half of the parallel sort as follows.

\begin{minted}{rust}
(input, buffer)
  .wrap_iter()
  .even_levels()
  .map(|(inp, out)| {
          inp.sort();
          (inp, out)
      }
  })
\end{minted}

This code just breaks the input and the buffer into small pieces
and sorts the input in-place (using the stable sequential sort from the Rust standard library)
once it can not be split any more. Note that the \emph{even\_levels}
adaptor will ensure data being merged back into the correct slice.

At this point different adaptors from Section~\ref{sec:control}
can be applied to control
the tasks divisions. We can use \emph{bound\_depth} to mimic
a classical divide and conquer, \emph{thief\_splitting} or
\emph{join\_context} to have a more dynamic splitting.
On top of that we can turn \emph{depjoin} (Section~\ref{sec:join}) on or off.
This gives us 6 different algorithms to try.

After choosing our adaptors, we continue with the
reductions which require merging sorted slices together.

We provide a generic \emph{merge} adaptor
fusing two sorted parallel iterators into one.
With this, we can easily implement the classical parallel merge of two sorted lists.
\begin{minted}{rust}
    reduce_with(|(l_in, l_out), (r_in, r_out)| {
      let output = fuse_slices(l_out, r_out);
          l_in.as_ref().into_par_iter()
              .merge(r_in.as_ref())
              .zip(output)
              .for_each(|(inp, out)| {
                  *out = *inp;
              });
      }
      (output, fuse_slices(l_in, r_in))
  });
\end{minted}

\mintinline{rust}{l_in.as_ref()} turns the left mutable slice as a slice
and \mintinline{rust}{into_par_iter} turns it into a parallel iterator on the left
elements.
The \emph{merge} adaptor uses by-default, an adaptive task splitting schedule (Section~\ref{sec:kaapi}) since divisions
come with a price (each division requires a binary search).
It is however possible to turn it back to a more classical
scheduling policy like \emph{thief\_splitting}.
After the merge, we just zip this iterator with a parallel iterator
on the elements of the output slice, and write the data into the buffer.

\paragraph{Merging slices instead of iterators}
We also provide a hand-tuned manual implementation
for an adaptive merge using the \emph{work} function directly
on slices (references of the vectors). The state is composed of the remaining input and
output slices. Working locally for $n$ iterations moves
the $n$ smallest elements from the two  input slices into the output slice.
This allows us to eliminate some bound checks that the iterators use internally.

The parallel sorts contain two levels of parallelism: parallelizing the sorts and each reduction.
Each stage can be tuned in many ways due to high number of adaptors we provide.
If we count here, we have 6 algorithms for the sorts and 3 different
reductions which means a total of 18 different parallel merge sorts.
Before \emph{Kvik} we ended up with the same algorithm re-implemented different times for
all the schedules that we wanted to try.

\begin{figure}[htbp]
    \begin{center}
        \includegraphics[width=7cm]{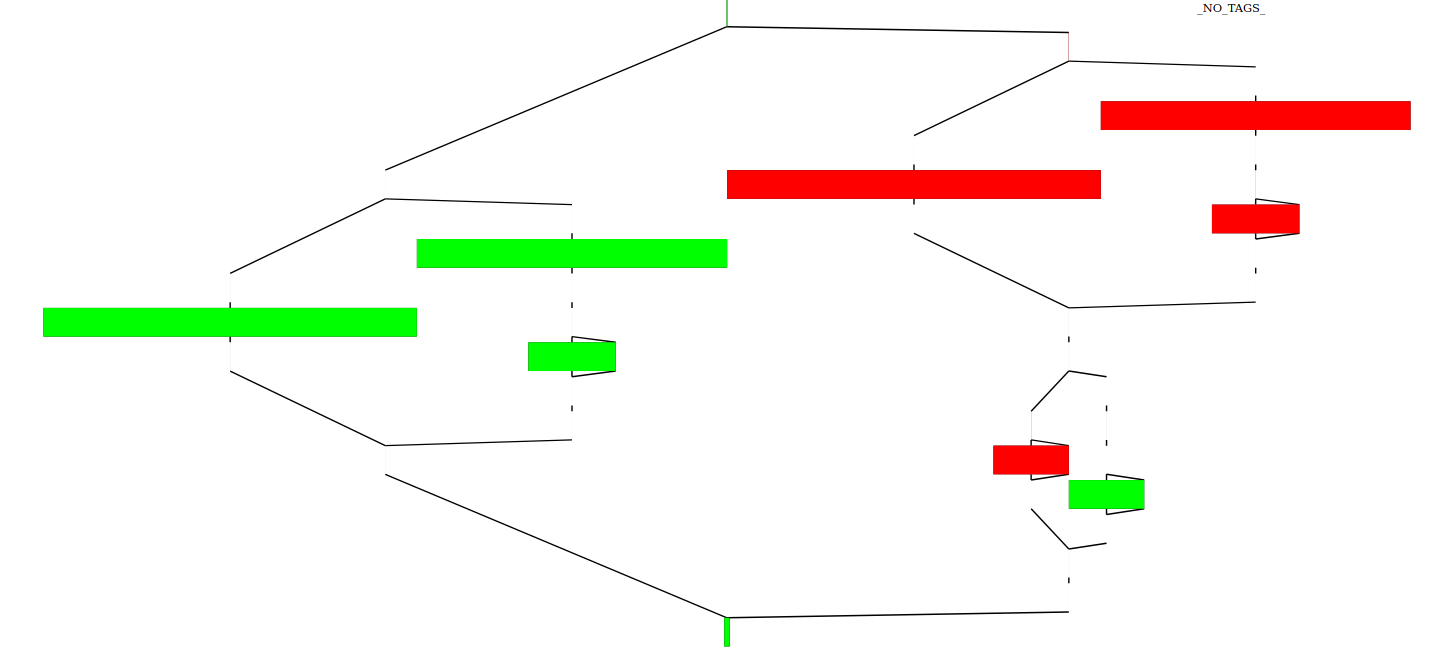}
    \end{center}
    \caption{Merge sort, two threads. The edges depict the splitting of tasks. For the sorting, there are a total of $4$ splits, two of which are in red and two in green. Directly underneath a pair of sorting tasks is a merge task. Finally, there is a merge that fuses the two sorted slices. Since this merge is larger, it got stolen and is split into two merge tasks in red and green.}
    \label{fig:sort}
\end{figure}

Figure~\ref{fig:sort} is the log obtained when running on two threads
with a depth limit of $2$ and the adaptive scheduling algorithm for the merge.

\section{Experiments}
\label{sec:expes}

We now proceed with a set of experiments comparing
\emph{Kvik} to state of the art libraries and different
scheduling policies between themselves. The goal
of this section is double: showing that choosing scheduling
policies matters and showing that \emph{Kvik} is competitive
with state of the art libraries.
We start Section~\ref{sec:exp_by_blocks} by testing
the \emph{by\_blocks} scheduler on very fine grain computations.
In Section~\ref{sec:exp_sort} we first compare different adaptors
before testing our stable sort algorithm against \emph{TBB} and
\emph{OpenMP}. Finally we conclude our experiments series in
Section~\ref{sec:pancakes} with an adaptive scheduling for
a benchmarks game~\cite{compbench} problem.

All the experiments for this work were conducted using 4 CPUs of Intel Xeon Gold 6130
(Skylake, 2.10GHz, 16 cores/CPU) with 768GiB total RAM, running Debian 10.5.
C++ programs were compiled using GCC version 8.3.0, C++17 standard and the \emph{O2}, -\emph{fopenmp} and \emph{-march=native} flags.
TBB 2020.3 was used for the sort benchmark, with \emph{ParallelSTL} header.
Rust programs were compiled with the \emph{rustc} compiler \emph{v1.45.0},
link time optimization enabled, and the \emph{target-cpu=native} flag.
In each experiment threads are always bound to cores in order to
minimize the number of numa nodes in use.
The lineplots for each experiment result include a confidence interval
with $95\%$ confidence level.

\subsection{Interruptible algorithms}
\label{sec:exp_by_blocks}

We start by a first set of experiments testing the \emph{by\_blocks} adaptor
from Section~\ref{sec:try_reduce}.
We try the blocking mechanism respectively on \emph{find\_first} and \emph{all}.
In both cases we consider an input vector of 100 million elements. In both cases,
sequential algorithms are blazing fast with sequential running times in the order of
tens of milliseconds. This shows that blocking mechanisms can be used even in fine grain
computations.

We compare the activation or de-activation of blocks and two different schedulers: \emph{thief\_splitting} and
\emph{adaptive}. The \emph{thief\_splitting} task splitting works as described Section~\ref{sec:TBB} and Section~\ref{sec:control}.
The adaptive scheduler works as detailed in Section~\ref{sec:adaptive} with an added benefit in this particular case.
Since the adaptive scheduler is regularly interrupting computations to check for steal requests we can take
advantage of this interruption to check if the task gets canceled due to the aborting mechanisms of our algorithms.
This allows us to stop any task as soon as it is recognized useless while the classical scheduler can only
cancel non started tasks.

In both sets of experiments we also compared ourselves with
implementations using the \emph{Rayon} library but for some reasons
the \emph{find\_first} implementation always ends in a speed-down, hence it is not shown in the curves.

\subsubsection{Find first}

\begin{figure}[htbp]
     \centering
     \def\svgwidth{\columnwidth}
     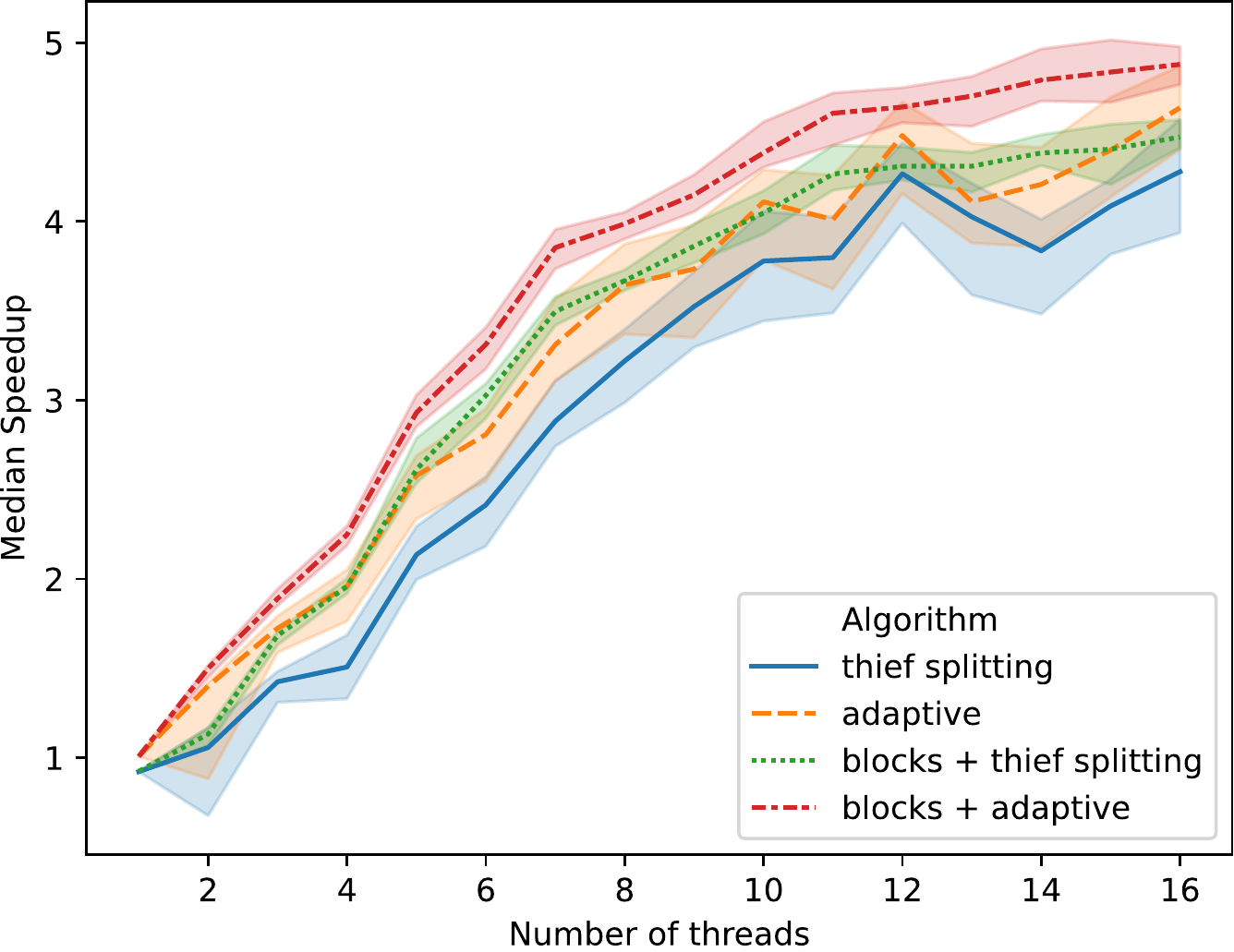
    \caption{Find-First speedups, uniformly distributed target. Note the high variability in the implementations without
    blocks.}
    \label{fig:find_random}
\end{figure}

\begin{figure}[htbp]
     \centering
     \def\svgwidth{\columnwidth}
     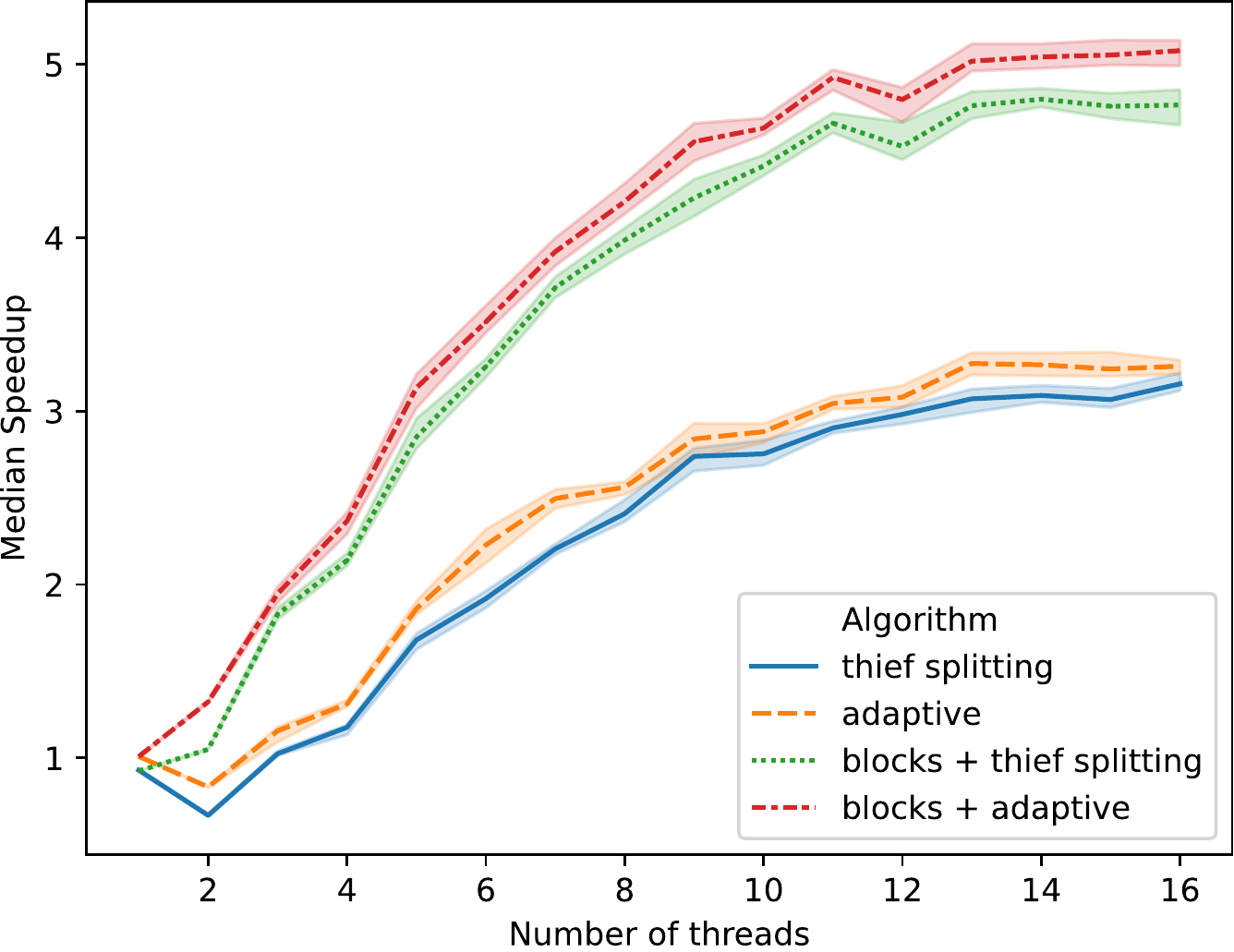
    \caption{Find-First speedups, target at $\frac{n}{2} -1$. The implementations without blocks slows down at $2$ threads since the first thread has to work until the end to reach the point of interruption. This is the point of maximum wasted work.}
    \label{fig:find_half}
\end{figure}

We start with \emph{find\_first}. Figure~\ref{fig:find_random} and Figure~\ref{fig:find_half} show
executions of our four versions of the find first algorithm on two different examples. In each case we only
have one target but its position differs. In Figure~\ref{fig:find_random} the position of the target is
chosen uniformly at random while Figure~\ref{fig:find_half} shows a target at $n/2-1$.
We can see that it is always better to activate blocks.
Implementations without blocks are much slower because, for any number of threads, at least half the work done is wasted.

\subsubsection{All}

\begin{figure}[H]
     \centering
     \def\svgwidth{\columnwidth}
     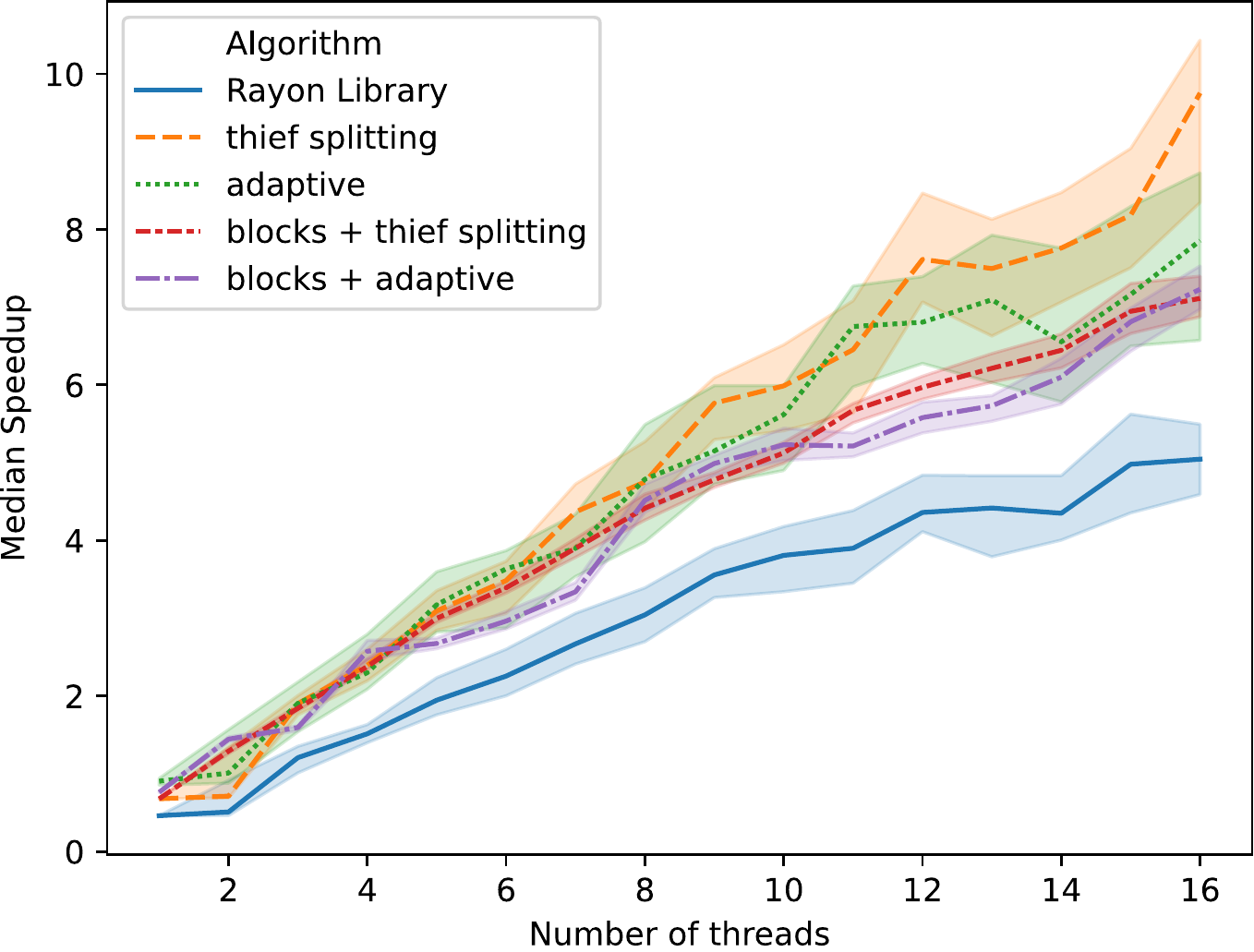
    \caption{Median speedups for the "All" with uniformly distributed target.
    Note the higher variability in speedups for implementations without blocks as opposed to the ones with blocks. The implementation from the \emph{Rayon} library is significantly slower.}
    \label{fig:all}
\end{figure}

Now for \emph{all} the measures are a bit more complex.
Figure~\ref{fig:all} shows the median speedups obtained.
A flexible implementation can be made with \emph{thief\_splitting} uses for task splitting.
Finally, the Adaptive scheduler (\ref{sec:kaapi}) can be used for minimal splits.
However, in this case there is no cost of division, so this policy does not yield a benefit.
Each of these policies can be composed with the \emph{by\_blocks} adaptor (Section~\ref{sec:try_reduce}) in order to add a sequential loop.
Implementations with blocks and without blocks have a similar performance but the confidence interval is much wider
for the variants without the blocks.
The variation in performance comes from the fact that the target may be found at the start of a task in the best case,
or at the end in the worst case.
Coupled with this, the other threads may have just started their tasks as well. They will all finish
their tasks before they can be stopped.
Since the blocks reduce the granularity of task splitting,
such different scenarios don't produce as much of a variation in time anymore.

\subsection{Parallel stable sort}
\label{sec:exp_sort}

All sorts benchmarks are done sorting a vector
containing random permutations of
of $10^8$ 32bits integers.
Speedups are obtained by comparing to the fastest
sequential algorithm which is in this case \emph{rust}'s
stable sort with a running time of $6.5$ seconds.

\subsubsection{Tuning the sort using adaptors}
\begin{figure}[H]
     \centering
     \def\svgwidth{\columnwidth}
     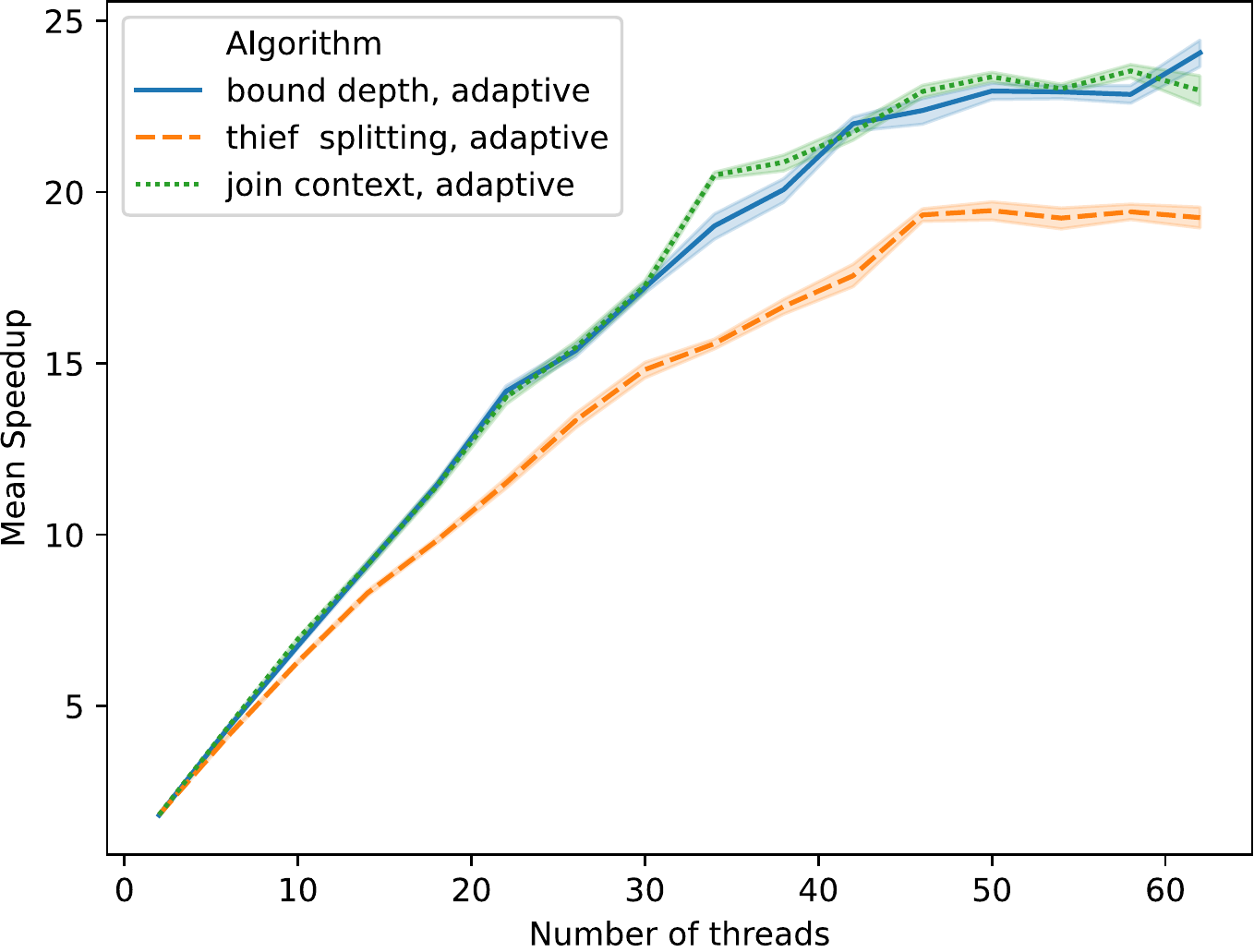
    \caption{A comparison of the same Iterator Sort with different adaptors for task splitting at the two levels. The name of each algorithm is given by the two adaptors it uses. Task splitting adaptors change the scalability of the same implementation.}
    \label{fig:sort_policy}
\end{figure}

We start by tuning the sort algorithm
from Section~\ref{sec:mergesort} using the task splitting adaptors.
We compare three adaptors: \emph{bound\_depth}, \emph{join\_context} and
\emph{thief\_splitting}. The \emph{thief\_splitting} policy doesn't need much calibration
but for the two others parameters are calibrated manually to their
best values.
The input to be sorted is first broken down into small pieces that are sorted in parallel,
then the sorted pieces are merged together, also in parallel. Hence, each curve in Figure~\ref{fig:sort_policy} has been obtained with a combination
of two adaptors from the list, one used for the sorting phase and the other for the merge.

We can see in this case that hand tuned policies win with
a slight advantage for the \emph{join\_context} adaptor.

The same process has been repeated for the \emph{depjoin}
adaptor and the different merging strategies but is not shown
here due to space constraints.
The best combination is the \emph{join\_context} composed with \emph{depjoin}
and the adaptive schedule for the merge.

\subsubsection{Comparing sorts}

We now compare our sorts with standard parallel stable
sorts algorithms from different libraries.

\begin{figure}
     \centering
     \def\svgwidth{\columnwidth}
     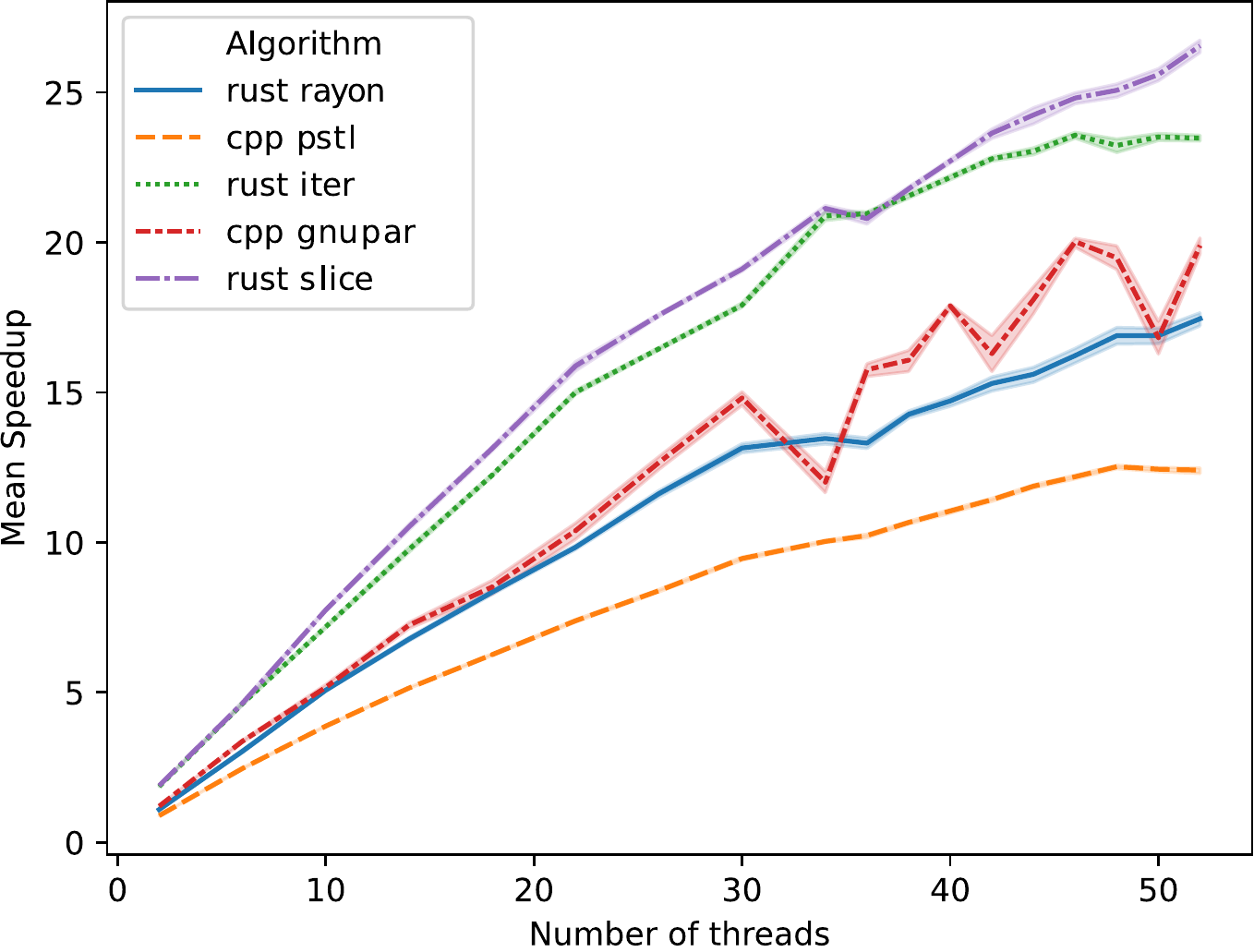
     \caption{Sort speedups across languages.}
     \label{fig:sort_bench}
\end{figure}
Figure~\ref{fig:sort_bench} shows the speedups obtained for different
threads number
 \emph{rust rayon} is the parallel stable sort from the Rayon library. \emph{cpp pstl} is the stable sort from the Intel ParallelSTL of C++ 17 with the \emph{par} execution policy. It internally links to TBB. \emph{cpp gnupar} is the GNU parallel stable sort using OpenMP. Finally we have the fastest stable sort, the slice sort as \emph{rust slice} and the Iterator sort as \emph{rust iter}

We can see a very good scaling for our algorithms with
speedups reaching up to 26, far more than speedups from
state of the art middlewares.
On top of that we can see that the more generic iterator based
merge algorithm does not degrade performances strongly when
compared to the manual slice based implementation.

One needs to be careful
about what conclusions to draw from this since we are based
on the fastest sequential algorithm. But at the very least
it shows we are competitive with state of the art libraries
even on very fine grain computations and all of this
with a sorting algorithm which is very clear and fits ten lines
of code.

\subsection{Fannkuch redux}
\label{sec:pancakes}
This micro benchmark has been described in \cite{fannkuch_redux}.
The problem is quite simple:

For all permutations of the sequence $(1..N)$, what is the maximum number of prefix-flips to be made in order to have $1$ at the start of the sequence.
A prefix-flip on a sequence $(j..N)$ is defined as reversal of the first $j$ elements of this sequence.
The program implementation has been taken from The Computer Language Benchmarks Game \cite{compbench}.

The implementation must (as per the rules of the benchmark) compute each permutation and then the flips required for it to reach the state $(1..)$. Hence, the complexity of this program is exponential in the input size. It is hence not memory-bound, since the sequence we work with is small ($length<16$).

Parallelism is exploited by dividing the set of all permutations of the sequence among the threads.
In the case of task-based implementations (for Rust), each task contains a set of permutations to work with.
This brings up an interesting caveat. Given a set of permutation-indices, generating the first permutation is much more
expensive as compared to generating the next permutation in a set from a given permutation of the set.
Therefore, task splitting becomes quite expensive, since the first permutation of the set assigned to the new task must be
generated from scratch.

We use $N=13$, and start off with the fastest code on the benchmark website (incidentally in Rust).
The implementation contains a parameter \emph{num\_blocks}, that indicates the number of sub-sets
that the set of all permutations
can be divided into. The implementation includes a parallel loop over all sub-sets (using the Rayon library).
Therefore, when the threads steal, they steal one or more sub-sets of permutations, and work on them.
For a competitive baseline, we do a grid search for \emph{num\_blocks} as a linear function of number of threads.
We find that a multiplier of $8$ is optimal for higher number of threads ($>40$). This baseline is named \emph{rust static}.

With \emph{Kvik}, we use the \mintinline{rust}{Work} (see Section~\ref{sec:work}).
The state in the \emph{work} preserves the permutation that a task had processed before being stolen.
One of the child-tasks can continue exactly where the parent left off, and quickly generate the next permutation.
Subsequently for the task splitting schedule, we can use the \emph{thief\_splitting} adaptor or better yet,
simply the Adaptive schedule.
These implementations are named \emph{rust thief} and \emph{rust adaptive} respectively.

\begin{figure}
     \centering
     \def\svgwidth{\columnwidth}
     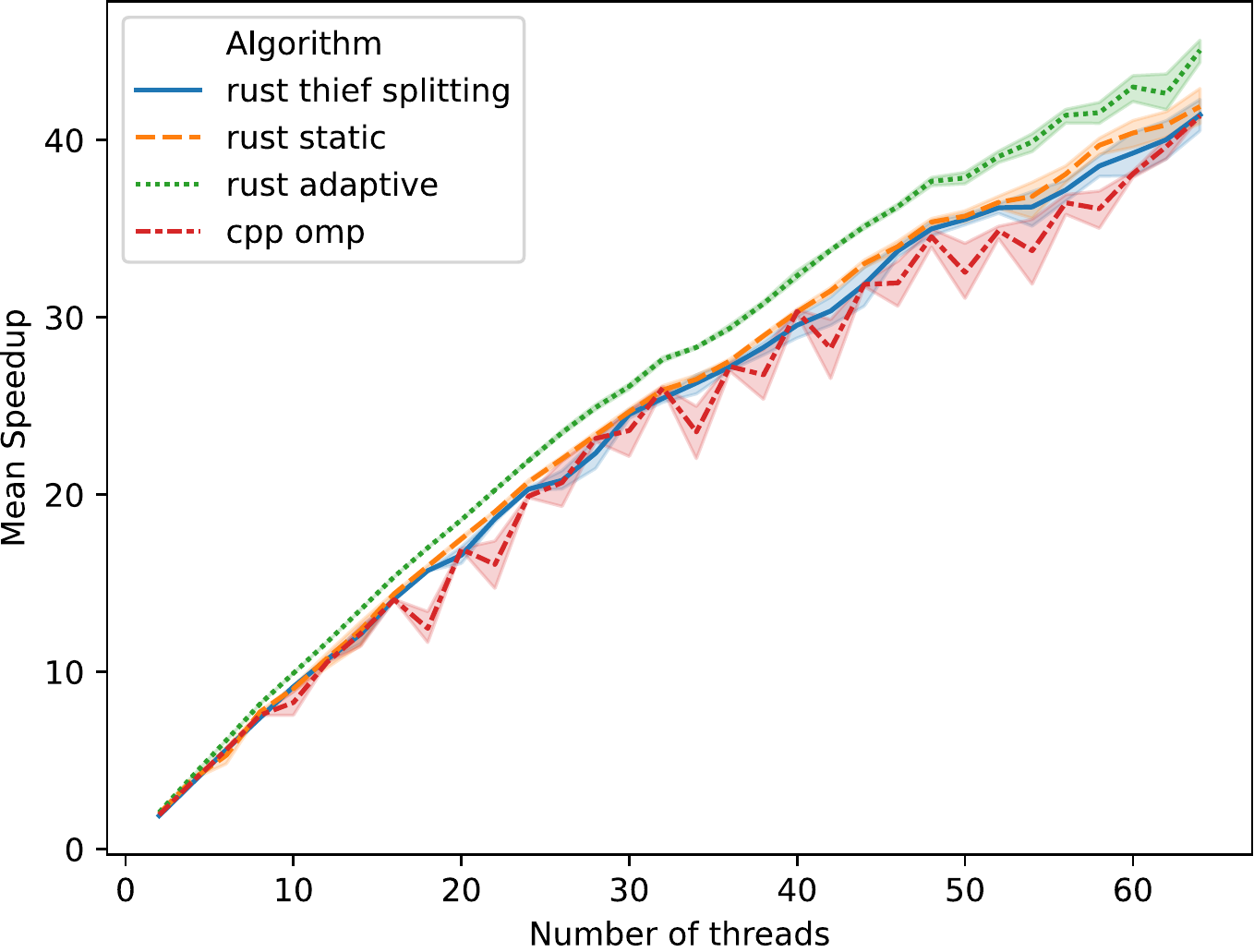
     \caption{The fannkuch redux benchmark.}
     \label{fig:pancakes_bench}
\end{figure}

Figure~\ref{fig:pancakes_bench} shows the comparison for these implementations along with the fastest C++ bench
taken as-is from the website, that has a simple \emph{omp parallel for}.
The frequent drops in speedup for this curve are also the regions of high variability (as shown by the shaded area).
This could be due to the slight load imbalance that manifests organically from the problem itself.
Since all rust implementations use task stealing as opposed to work sharing, they exhibit stable performances.
Among the rust implementations, adaptive schedule leads due to the minimal task splits.
The tuned \emph{static} implementation from the website that uses the Rayon library,
is quite close to the one with \emph{thief\_splitting} implemented in \emph{Kvik}. This means that we can achieve the same performance
without the need for tuning extra parameters.

\section{Conclusion}
\label{sec:conclusion}

In this paper we presented our work on \emph{Kvik}, a prototype task-based middleware built
on top of \emph{Rayon}. \emph{Kvik} allows end-users to finely tune their algorithms' behavior
through a mix of task splitting schedulers and adaptors that control the splits themselves. Out of these, the \emph{by\_blocks}
adaptor allows to add an external sequential loop which proves itself valuable for interruptible
computations.
We also provide a nice abstraction
for divide and conquer algorithms allowing to delegate tasks creations decisions to the middleware.
In our opinion the \emph{composability} we put at the disposition of parallel programmers is unmatched, allowing
simple and elegant code.

Among the schedulers provided, the adaptive scheduler can be used to reduce tasks creations to a bare minimum, and balance the load organically.
It allows iterators to be consumed gradually while keeping sequential executions fast. We demonstrate its superiority
through different experiments. Early interruptions in the \emph{find} and \emph{all} algorithms,
adaptive merges in our sort algorithms and adaptive divisions for the pancakes benchmark, all result in a fast and scalable implementation.

Our experiments show that is possible to provide genericity while keeping strong performances as is
demonstrated by a fast and scalable iterator sort implementation.

In our future works we hope to work our way towards
an integration within the \emph{Rayon} library. We also take interest in the composability of parallel
algorithms. We also expect to work on locality issues.
\emph{By\_blocks} is already having nice properties for locality
but it could be improved in different ways. We also consider enhancing
\emph{Divisible} to execute different kind of divisions
based on stealers ids.
Finally we hope to tackle \emph{streams} and \emph{futures}.
Interruptible parallel iterators
will allow a mix of long computations interrupted to allow fast IO.

\begin{acks}
\section{Acknowledgements}
Experiments presented in this paper were carried out using the Grid'5000 testbed, supported by a scientific interest group hosted by Inria and including CNRS, RENATER and several Universities as well as other organizations (see \href{https://www.grid5000.fr}{grid5000.fr})\cite{grid5000}.
We also thank Bruno Raffin for his invaluable contributions to the orgranisation of the paper.
\end{acks}

\bibliographystyle{ACM-Reference-Format}
\bibliography{ppopp.bib}


\end{document}

%% file: ffrand.pdf_tex
\begingroup%
  \makeatletter%
  \providecommand\color[2][]{%
    \errmessage{(Inkscape) Color is used for the text in Inkscape, but the package 'color.sty' is not loaded}%
    \renewcommand\color[2][]{}%
  }%
  \providecommand\transparent[1]{%
    \errmessage{(Inkscape) Transparency is used (non-zero) for the text in Inkscape, but the package 'transparent.sty' is not loaded}%
    \renewcommand\transparent[1]{}%
  }%
  \providecommand\rotatebox[2]{#2}%
  \newcommand*\fsize{\dimexpr\f@size pt\relax}%
  \newcommand*\lineheight[1]{\fontsize{\fsize}{#1\fsize}\selectfont}%
  \ifx\svgwidth\undefined%
    \setlength{\unitlength}{384.56063474bp}%
    \ifx\svgscale\undefined%
      \relax%
    \else%
      \setlength{\unitlength}{\unitlength * \real{\svgscale}}%
    \fi%
  \else%
    \setlength{\unitlength}{\svgwidth}%
  \fi%
  \global\let\svgwidth\undefined%
  \global\let\svgscale\undefined%
  \makeatother%
  \begin{picture}(1,0.77196736)%
    \lineheight{1}%
    \setlength\tabcolsep{0pt}%
    \put(0,0){\includegraphics[width=\unitlength,page=1]{ffrand.pdf}}%
  \end{picture}%
\endgroup%

%% file: ffmid.pdf_tex
\begingroup%
  \makeatletter%
  \providecommand\color[2][]{%
    \errmessage{(Inkscape) Color is used for the text in Inkscape, but the package 'color.sty' is not loaded}%
    \renewcommand\color[2][]{}%
  }%
  \providecommand\transparent[1]{%
    \errmessage{(Inkscape) Transparency is used (non-zero) for the text in Inkscape, but the package 'transparent.sty' is not loaded}%
    \renewcommand\transparent[1]{}%
  }%
  \providecommand\rotatebox[2]{#2}%
  \newcommand*\fsize{\dimexpr\f@size pt\relax}%
  \newcommand*\lineheight[1]{\fontsize{\fsize}{#1\fsize}\selectfont}%
  \ifx\svgwidth\undefined%
    \setlength{\unitlength}{384.56063474bp}%
    \ifx\svgscale\undefined%
      \relax%
    \else%
      \setlength{\unitlength}{\unitlength * \real{\svgscale}}%
    \fi%
  \else%
    \setlength{\unitlength}{\svgwidth}%
  \fi%
  \global\let\svgwidth\undefined%
  \global\let\svgscale\undefined%
  \makeatother%
  \begin{picture}(1,0.77196736)%
    \lineheight{1}%
    \setlength\tabcolsep{0pt}%
    \put(0,0){\includegraphics[width=\unitlength,page=1]{ffmid.pdf}}%
  \end{picture}%
\endgroup%

%% file: all.pdf_tex
\begingroup%
  \makeatletter%
  \providecommand\color[2][]{%
    \errmessage{(Inkscape) Color is used for the text in Inkscape, but the package 'color.sty' is not loaded}%
    \renewcommand\color[2][]{}%
  }%
  \providecommand\transparent[1]{%
    \errmessage{(Inkscape) Transparency is used (non-zero) for the text in Inkscape, but the package 'transparent.sty' is not loaded}%
    \renewcommand\transparent[1]{}%
  }%
  \providecommand\rotatebox[2]{#2}%
  \newcommand*\fsize{\dimexpr\f@size pt\relax}%
  \newcommand*\lineheight[1]{\fontsize{\fsize}{#1\fsize}\selectfont}%
  \ifx\svgwidth\undefined%
    \setlength{\unitlength}{390.9231369bp}%
    \ifx\svgscale\undefined%
      \relax%
    \else%
      \setlength{\unitlength}{\unitlength * \real{\svgscale}}%
    \fi%
  \else%
    \setlength{\unitlength}{\svgwidth}%
  \fi%
  \global\let\svgwidth\undefined%
  \global\let\svgscale\undefined%
  \makeatother%
  \begin{picture}(1,0.75940314)%
    \lineheight{1}%
    \setlength\tabcolsep{0pt}%
    \put(0,0){\includegraphics[width=\unitlength,page=1]{all.pdf}}%
  \end{picture}%
\endgroup%

%% file: sort_policy.pdf_tex
\begingroup%
  \makeatletter%
  \providecommand\color[2][]{%
    \errmessage{(Inkscape) Color is used for the text in Inkscape, but the package 'color.sty' is not loaded}%
    \renewcommand\color[2][]{}%
  }%
  \providecommand\transparent[1]{%
    \errmessage{(Inkscape) Transparency is used (non-zero) for the text in Inkscape, but the package 'transparent.sty' is not loaded}%
    \renewcommand\transparent[1]{}%
  }%
  \providecommand\rotatebox[2]{#2}%
  \newcommand*\fsize{\dimexpr\f@size pt\relax}%
  \newcommand*\lineheight[1]{\fontsize{\fsize}{#1\fsize}\selectfont}%
  \ifx\svgwidth\undefined%
    \setlength{\unitlength}{390.9231369bp}%
    \ifx\svgscale\undefined%
      \relax%
    \else%
      \setlength{\unitlength}{\unitlength * \real{\svgscale}}%
    \fi%
  \else%
    \setlength{\unitlength}{\svgwidth}%
  \fi%
  \global\let\svgwidth\undefined%
  \global\let\svgscale\undefined%
  \makeatother%
  \begin{picture}(1,0.75940314)%
    \lineheight{1}%
    \setlength\tabcolsep{0pt}%
    \put(0,0){\includegraphics[width=\unitlength,page=1]{sort_policy.pdf}}%
  \end{picture}%
\endgroup%

%% file: sort.pdf_tex
\begingroup%
  \makeatletter%
  \providecommand\color[2][]{%
    \errmessage{(Inkscape) Color is used for the text in Inkscape, but the package 'color.sty' is not loaded}%
    \renewcommand\color[2][]{}%
  }%
  \providecommand\transparent[1]{%
    \errmessage{(Inkscape) Transparency is used (non-zero) for the text in Inkscape, but the package 'transparent.sty' is not loaded}%
    \renewcommand\transparent[1]{}%
  }%
  \providecommand\rotatebox[2]{#2}%
  \newcommand*\fsize{\dimexpr\f@size pt\relax}%
  \newcommand*\lineheight[1]{\fontsize{\fsize}{#1\fsize}\selectfont}%
  \ifx\svgwidth\undefined%
    \setlength{\unitlength}{390.9231369bp}%
    \ifx\svgscale\undefined%
      \relax%
    \else%
      \setlength{\unitlength}{\unitlength * \real{\svgscale}}%
    \fi%
  \else%
    \setlength{\unitlength}{\svgwidth}%
  \fi%
  \global\let\svgwidth\undefined%
  \global\let\svgscale\undefined%
  \makeatother%
  \begin{picture}(1,0.75940314)%
    \lineheight{1}%
    \setlength\tabcolsep{0pt}%
    \put(0,0){\includegraphics[width=\unitlength,page=1]{sort.pdf}}%
  \end{picture}%
\endgroup%

%% file: fannkuch_redux.pdf_tex
\begingroup%
  \makeatletter%
  \providecommand\color[2][]{%
    \errmessage{(Inkscape) Color is used for the text in Inkscape, but the package 'color.sty' is not loaded}%
    \renewcommand\color[2][]{}%
  }%
  \providecommand\transparent[1]{%
    \errmessage{(Inkscape) Transparency is used (non-zero) for the text in Inkscape, but the package 'transparent.sty' is not loaded}%
    \renewcommand\transparent[1]{}%
  }%
  \providecommand\rotatebox[2]{#2}%
  \newcommand*\fsize{\dimexpr\f@size pt\relax}%
  \newcommand*\lineheight[1]{\fontsize{\fsize}{#1\fsize}\selectfont}%
  \ifx\svgwidth\undefined%
    \setlength{\unitlength}{390.9231369bp}%
    \ifx\svgscale\undefined%
      \relax%
    \else%
      \setlength{\unitlength}{\unitlength * \real{\svgscale}}%
    \fi%
  \else%
    \setlength{\unitlength}{\svgwidth}%
  \fi%
  \global\let\svgwidth\undefined%
  \global\let\svgscale\undefined%
  \makeatother%
  \begin{picture}(1,0.75940314)%
    \lineheight{1}%
    \setlength\tabcolsep{0pt}%
    \put(0,0){\includegraphics[width=\unitlength,page=1]{fannkuch_redux.pdf}}%
  \end{picture}%
\endgroup%